\documentclass[useAMS,usenatbib]{mn2e} 
\usepackage{graphicx} 
\usepackage{txfonts} 
\usepackage{graphicx}
\usepackage{color,epsfig,amssymb,longtable}
\usepackage{array,colortbl,lscape}

\bibpunct[,]{(}{)}{;}{a}{,}{,}

\begin{document}

\title[Modelling the evolution of H{\sc ii} galaxies]{The evolution of
H{\sc ii} galaxies: Testing the bursting scenario through the use of
self-consistent models}

\author[{Mart\'{\i}n-Manj\'{o}n} et al.]  {M.L. Mart\'{\i}n-Manj\'{o}n
$^{1}$ \thanks{E-mail:mariluz.martin@uam.es}, M.~Moll\'{a},$^{2}$,
A.~I.~D\'{\i}az$^{1,3}$, and R.~Terlevich$^{3}$ \thanks{Research Affiliate 
IoA, Cambridge} \\ $^{1}$Departamento
de F\'{\i}sica Te\'{o}rica, Universidad Aut\'onoma de Madrid, 28049
Cantoblanco, Madrid (Spain)\\ $^{2}$Departamento de Investigaci\'{o}n
B\'{a}sica, CIEMAT, Avda. Complutense 22, 28040, Madrid, (Spain) \\
$^{3}$ INAOE, Luis Enrique Erro 1, Tonanzintla, Puebla 72840, Mexico}


\date{Accepted Received ; in original form }

\pagerange{\pageref{firstpage}--\pageref{lastpage}} \pubyear{2006}

\maketitle \label{firstpage}

\begin{abstract} 

We have computed a series of realistic and
self-consistent models of the emitted spectra of H{\sc ii} galaxies.
Our models combine different codes of chemical evolution, evolutionary
population synthesis and photoionization. The emitted spectrum of
H{\sc ii} galaxies is reproduced by means of the photoionization code
CLOUDY, using as ionizing spectrum the spectral energy distribution of
the modelled H{\sc ii} galaxy, which in turn is calculated according
to a Star Formation History (SFH) and a metallicity evolution given by
a chemical evolution model that follows the abundances of 15 different
elements. The contribution of emission lines to the broad-band colours
is explicitly taken into account.

The results of our code are compared with photometric and
spectroscopic data of H{\sc ii} galaxies. Our technique reproduces
observed diagnostic diagrams, abundances, equivalent width-colour and
equivalent width-metallicity relations for local H{\sc ii} galaxies.

\end{abstract}


\section{Introduction}

H{\sc ii} galaxies, a subset of Blue Compact Galaxies (BCG), are
 gas rich dwarf galaxies whose optical spectra are dominated by strong
 and narrow emission lines. They are currently experiencing intense
 star formation in small volumes. The emission lines are produced by
 gas ionized by the young massive stars.  Observations indicate that
 H{\sc ii} galaxies are in general metal poor systems.  This fact in
 addition to their very young stellar populations has been known for
 some time and has led to the proposal that these systems are very
 young, suffering their first burst of star formation \citep{SS70}.

It seems, however, that these galaxies are not as young as it was
originally thought, since recent photometric observations indicate that
they host stellar populations which are at least $10^{7}-10^{8}$ a. old,
reaching in some cases a few Ga
\citep{tel97,leg00,tols03,cai03,vzee04,thuan05}.
There is now wide
agreement about the existence of underlying populations with ages of
several Gas in BCGs and/or H{\sc ii} galaxies.
Even in the most metal-poor galaxy known, Izw18, the best candidate
to {\sl primordial galaxy}, there is an underlying population of at
least some 10$^{8}$--10$^{9}$ a \citep{izo04}.

However there is no consensus about their history of star formation.
Two scenarios are postulated:

a) The star formation takes place in short but intense episodes separated
by long quiescent periods of null or low activity \citep{Bra98}.

b) The star formation is continuous and of low intensity along the galaxy
life, with superimposed sporadic bursts \citep {leg00}.

In both scenarios, a stellar population of intermediate age
contributes to the observed luminosity. Indeed, recent work takes into
account the existence of an underlying stellar population and combine
photo-ionization codes with evolutionary synthesis models (hereinafter
ESM).  Examples of this procedure can be found in \citet{gar99},
\citet{mas99}, \citet{sta03} and \citet{moy01}.  In general the ESM
so far do not include the chemical evolution.
Given the low metal content of H{\sc ii} galaxies the metallicity effect can
be of considerable relevance.

\citet{ter04} have shown that the equivalent width of the H$\beta$
line, EW(H$\beta$), which can be taken as an age indicator for the
ionizing population, and the (U-V) colour in H{\sc ii} galaxies show
an anti-correlation which cannot be interpreted in terms of a single
stellar population (SSP).  At a given EW(H$\beta$) the data are
displaced to colours redder than those predicted by the models. The
most likely explanation is that the ionizing population, that is the
most recent burst of star formation, is overimposed on an older
stellar population evolved enough as to produce a redder (U-V) colour.
There is also an inverse correlation between EW(H$\beta$) and the gas
oxygen abundance, $12+log(O/H)$, which could be interpreted in terms
of galactic chemical evolution: as the average age and the total mass
of the stellar population increases , the EW(H$\beta$) decreases and
the mean metallicity of the interstellar medium increases.

A number of works \citep[and references therein]{shi06}
have computed purely chemical evolution models (hereinafter CEM) for
the study of BCG and dwarf irregular galaxies. Most of them assume that
the star formation occurs in bursts and include the effects of galactic
winds and/or gas infall. However most of them limit the study to the
evolution of nitrogen and oxygen abundances \citep{HEK00,lar01} and/or the
luminosity-metallicity relation \citep{mouh02}. Only \citet{vaz03} used the
information coming from the chemical evolution models from \citet{car02}
to perform the next step and combine chemical and spectral evolution.
Their models exclude however the early stages of evolution, i.e. during
the nebular stage when the most massive stars dominate the energy output.

A code combining chemical evolution, evolutionary synthesis and
photo-ionization models has not yet been developed and used for the
spectral analysis of BCG and H{\sc ii} galaxies.

In this work, we  include the chemical evolution model results
in the computation of the spectral energy distributions which are
used as ionizing source for a photo-ionization code.

Our method allows the simultaneous use of the whole available
information for the galaxy sample concerning on the one hand the
ionized gas -- emission lines intensities and equivalent widths,
elemental abundances, gas densities -- which defines the present time
state of the galaxy, and on the other hand spectro-photometric
parameters -- colours, absorption spectral indices --- defined by the
stellar populations and their evolution with time and thus related to
the galaxy star formation history. This is done in a self-consistent
way, that is using the same assumptions regarding stellar evolution,
model stellar atmospheres and nucleosynthesis, and using a realistic
age-metallicity relation.

The basic codes employed are described in Section 2. Section 3 shows
the corresponding results of each kind of model, and the
whole combined results about galaxy evolution. These results are discussed 
in Section 4.  Finally, our conclusions are summarized in Section 5.

\section{Model description}

\subsection{Chemical evolution}

The code used for this work is a generalization of the code used for the
Solar Neighborhood by \citet{fer92} and applied to the whole Galaxy disc
in \citet{fer94}. That model described the galaxy as a two-zone system (halo
and disc) in which the disc is a secondary structure formed by the
gravitational accumulation of gas from the halo.  In the present work, 
however,  all the gas is assumed to be within an only region from $t=0$, 
that is, the collapse time scale has been eliminated as an input parameter 
of the code. The matter is assumed to be in different phases: 

\begin{itemize} 
\item[-]A stellar population, where we distinguish the stars able to create 
and eject heavy elements to the interstellar medium  ($M > 4M_{\odot}$), 
and the stars  which only eject hydrogen and He ($M < 4M_{\odot}$).  
\item[-]Stellar remnants, as the endpoint of star evolution, that act as a 
matter sink, removing mass from chemical evolution.  
\item[-]Interstellar diffuse gas out of which stars are forming 
following a simple Schmidt law. No molecular component is explicitly 
considered in the present work.  
\end{itemize}

A bursting star formation has been assumed taking place 
in successive bursts along the time evolution. At any time, the available
gas for each burst is the sum of the gas left after the previous burst
of star formation and the gas ejected by massive stars.

According to this description, the time dependence of the total mass
fraction in each phase and the chemical abundances in the ISM are
determined by the interactions between these phases and, as a consequence
of the evolution, we get the star formation rate. The basic equations
to study the behavior of the mass of stars and gas are, therefore:
\begin{eqnarray} 
\frac{dM}{dt}& = & 0 \\ 
\frac{dM_{s}}{dt}& = &  \Psi - E \\ 
\frac{dM_{g}}{dt}& = & -\Psi + E \\
M & = & M_{s}+M_{g}
\end{eqnarray}
where $\rm M_{s}(t)$ is the mass in stars, M$_{g}$(t) is the mass of
gas, $M$ is the total mass of the system, $\Psi$(t) the star formation
rate and E(t) the ejection rate of mass from the stars to the ISM. A
part of the restituted gas consists of enriched material, E$_{Z}$(t),
which is created in the interior of the stars and is ejected when they
die. To compute both quantities, ejected gas and element production,
we use the {\sl $Q_{ij}$ matrix} formalism introduced by \citet{tal73}
and widely used in classical chemical evolution models. It is very
useful to treat elemental abundances which increase at different rate
in the interstellar medium, that is, the non-solar elemental ratios
\citep{pcb98}. In the present work we adopt the same matrix
prescription as in \citet{gav06}, which is an updated version from the
one given in \citet{fer92} and \citet{gal95} calculated following the
method described by \citet{pcb98}.

In order to calculate these matrices it is necessary
to consider the nucleosynthesis of stars, their initial mass function (IMF) and
their end either as type I or II supernovae, or through quiet evolution.
The mean stellar lifetimes, final states, and nucleosynthesis products,
depend on the stellar mass which makes the IMF to play an important role
in the chemical evolution model.  The adopted IMF is assumed to be universal
in space and constant in time \citep{wys97,sca98,mey00} and taken
from \citet{fer90}. This IMF is very similar to Scalo's law \citep{sca86}
and in good agreement with the recent expressions from \citet{kro01}
and \citet{cha03}.  Nucleosynthesis yields for massive stars have been
taken from \citet{woo95}. For low and intermediate mass stars, we have used
the yields from \citet{gav05}.
The combination of these sets of stellar yields, with our
assumed IMF producing the required yields per stellar generation, has
been completely revised and calibrated with the MWG in \citet{gav05,
gav06}, where other stellar yields sets have been used and analyzed
too. It also reproduces observations of spiral and irregular galaxies
\citep{mol05}, obtaining reasonable results even for nitrogen
abundances in low-metallicity objects \citep{mol06}.
For type I supernova explosion releases,
model W7 from \citet{nom84}, as revised by \citet{iwa99}, has been taken,
 while the rate of this type of SNIa
explosions are included through a table given by  Ruiz-Lapuente (private
communication), following works of \citet{ruiz00}. 

We have assumed that the gas is used to form stars with a given efficiency
called H: $\Psi=H\cdot M_{g} $. If there would not be gas
ejection from massive stars, as it would be the case for instance
in the firsts moments of the evolution, $(E=0)$, then:

\begin{eqnarray} 
\frac{dM_{s}}{dt}& = &  HM_{g} \\
\frac{dM_{g}}{dt} &= & -HM_{g} 
\end{eqnarray}

That is, the consumed gas rate would be a function that depends on
the available gas at a given time, \footnote{When $E\ne 0$
this expression is modified, as  we will show in Section 3.} 
being a decreasing function
of time through the parameter H, which defines the star formation
efficiency. Then:
 \begin{equation} 
\frac{dM_{g}}{M_{g}}=-Hdt \Rightarrow M_{g}(t)=M_{g0}e^{-Ht} 
\end{equation}

A change in the star formation rate implies a change in the
value of the parameter H or efficiency: 
\begin{equation}
H=\frac{ln\frac{M_{g,0}}{M_{g}}}{\Delta t} 
\end{equation}

We have run models considering the star formation as a set of
successive bursts in a region with a total mass of gas of
100$\cdot$10$^{6}$ M$_{\odot}$ and a radius of 500 pc (see section 3.2
for details).  Every galaxy experiences 11 star formation bursts along
its evolution of 13.2 Ga, that is one every 1.3 Ga.  In each burst a
certain amount of gas is consumed to form stars. We have computed two
types of models. In the first one, which we call {\sl burst} models
(models B), the same efficiency H is assumed for every burst while in
the second one (models A) we assume an attenuated bursting star
formation mode , {\em i. e.} a variable H is assumed along the time
evolution.  For each type we have computed two models with different
initial star formation efficiency:
\begin{enumerate}
\item[-]model 1, that uses $\sim$ 64 per cent of the gas for star formation in
each time step. Thus, following Eq.(8), $1/H =0.5$ Ma
\item[-]model 2, that uses $\sim$ 33 per cent of the gas to form stars in each
time step, which implies $1/H=1.25$ Ma 
\end{enumerate} 
These percentages are maintained in models B, which means that every
burst will transform the corresponding percentage of the available gas
(taking into account the gas ejected by massive stars) in stars. In
Models A, these values correspond to the initial efficiencies and are
reduced in the successive bursts by a factor $n$, where $n$ is the number
of the burst.

The code solves the chemical evolution equations to obtain, in each
time step, the abundances of 15 elements: H, D, $\rm ^{3}He$, $\rm
^{4}He$, C, $\rm ^{13}C$, O, N, Ne, Mg, Si, S, Ca, Fe, and {\it nr} where 
{\it nr}
are the isotopes of the neutron rich elements, synthesized from $\rm
^{12}C$, $\rm ^{13}C$, $^{14}$N and $\rm ^{16}O$ inside the CO
core. We have taken a time step, $\Delta$t= 0.5 Ma \footnote{This
small time step allows to take into account the fast evolutionary
phases of the most massive stars considered.}, from the initial time,
$t=0$, up to the final one, $t= 13.2$ Ga.
Also, at each time step, the star formation rate and the mass in each
phase -- low mass, massive stars and remnants, total mass in stars
created, and mass of gas-- are computed. This amounts to a total of 26348
stellar {\sl generations} or SSPs, each one with its corresponding age
and metallicity.

\subsection{Evolutionary synthesis}

Once the chemical evolution models have been calculated, the
spectrophotometric properties can also be obtained.  In this work we
have used the code from \citet{garc95a,garc95b,gmb98} as updated in
\citet{mol00}.  The required inputs are the stellar isochrones and the
stellar atmosphere models for the individual stars. In that work we
used the set of 50 isochrones from the Padova group from \citet{ber94}
for ages between 4 Ma and 20 Ga, for 5 metallicities between 1/50
and 2.5 Z$_{\odot}$ (Z$=$0.0004, 0.004, 0.008, 0.02 and 0.05).  To this
set of isochrones we have added other 18 isochrones from \citet{bre93} 
and \citet{fag94} with ages from 0.5 to 4 Ma, used in \citet{garc95a} 
and \citet{gmb98} for the youngest stellar populations of 4 metallicities 
Z$=$0.004, 0.008, 0.02 and 0.05. That means that  for the lowest metallicity,
$Z=0.0004$, only isochrones older than 4 Ma are available.

 In both cases, isochrones give the number of stars in
each phase assuming that stellar clusters were formed with a standard
Salpeter IMF with $m_{low}=0.6$ M$_{\odot}$ and m$_{up}=100$
M$_{\odot}$. This IMF, although using the same lower and upper limits,
is not exactly the one used in the multiphase chemical evolution code,
what implies some differences in the resulting spectra, mainly for the
oldest and the youngest stellar populations. They are, however,
smaller than 10 per cent (Garc\'{\i}a-Vargas et al. 2008, in preparation),
and therefore we assume that this inconsistency will not produce very
important differences in the final results. 

The emergent spectral energy distribution (SED) was synthesized by
calculating the number of stars in each point of the H-R diagram and
assigning to it the most adequate stellar atmosphere model. The models
of \citet{clegg87} and \citet{lej97} have been used for stars with
$T_{eff} \geq$ 50000 K (last evolutionary stages of massive stars) and
2500 $\leq T_{eff}< 50000$ K, respectively.
  
To each SED computed for a stellar cluster, it was added a continuum
nebular emission contribution as explained in \citet{gmb98}. The gas is
assumed to have an electron temperature, T$_{e}$ which depends on Z.
The assumed values are T$_{e}=$18000, 11000, 9000, 6500 and 4000 K for
Z$=$0.0004, 0.004, 0.008, 0.02 and 0.05, respectively, selected according to
the average value obtained by \citet{garc95b}. The free-free, free-bound 
emission by hydrogen and neutral helium, and the two photon hydrogen-continuum
have been calculated by means of the atomic data from \citet{all84} and
\citet{fer80}. 

The SEDs for the complete sets of 68 (50 for Z=0.0004) isochrones were
calculated with the same code as in \citet{mol00}
\footnote{Although in that work, applied to old elliptical galaxies,
the SEDs corresponding to ages younger than 4 Ma and those for
Z$=$0.0004, were not used} for the cited metallicities.  Therefore we
finally have a set of 68 SEDs for SSP of different ages for 4
metallicities each, and 50 for ages older than 4 Ma for Z$=$0.0004,
which we use as a spectral library.

Once the SEDs for SSPs are computed, they must be convolved with the star 
formation history (SFH) in order to calculate the corresponding SED 
for a region where more than a burst take place:
\begin{equation}
 L_{\lambda}(t)=\int^{t}_{0}S_{\lambda}(\tau,Z(t'))\Psi(t')dt'
\label{sp}
\end{equation}
where $\tau=t-t'$ is the age of the stellar population created in a time
$t'$ and $S_{\lambda}$ being the SED for each SSP of age
$\tau$ and metallicity Z reached in that time $t'$.
A SED from the SSP library, $S_{\lambda}$, must assigned to each time step according to
its corresponding age and metallicity taking into account the SFH,
$\Psi(t)$, and the AMR, $Z(t)$, obtained from the chemical evolution
model, to finally calculate L$_{\lambda}(t)$ by the above integration.  
However the metallicity changes continuously  while the available SEDs have
only 4 or 5 possible values. We have, therefore, interpolated logarithmically
between the two SSP of the same age $\tau$ closest in metallicities 
to Z(t') to obtain the corresponding $S_{\lambda}(\tau,Z(t'))$.
The final result is the total luminosity at each wavelength $\lambda$. 

When this process is applied to the stellar population created by the
first burst two problem arise. On the one hand the initial metallicity is
Z$=0$ and, although during the first Ma after the creation of the
first stars the metallicity increases, it does not reach the minimum
Z$=$0.0004, and therefore we must extrapolate with the available SEDs
of the youngest SSPs.  Due to the uncertainties in the evolution of the
{\sl quasi}-zero-metallicity stars, we have preferred to assign to this
first burst metallicity a minimum value of 0.0028, smaller than the
one reached at the end of the first burst but still valid for
extrapolating without creating instability numerical problems. On the other
hand we must to use two different pairs of metallicities to extrapolate
with the SEDs of SSPs: Z$=$0.004 and 0.008 before 4 Ma and Z$=$0.0004
and 0.004 after 4 Ma, since SEDs for the most-metal-poor
and youngest SSPs are not available. 
We have checked that this method does not produce
discontinuities and that the resulting spectra behave smoothly.
In this way we obtain the SED
corresponding to the whole stellar population, including the ionizing
continuum proceeding from the last formed stellar population.

\subsection{Photo-ionization}

The photo-ionization code CLOUDY, in its 96.0 version \citep{fer98},
has been used in order to obtain the emission lines. The gas is
assumed to be ionized by the massive stars belonging to the current
burst of star formation whose SED has been previously calculated by
the combination of the chemical and evolutionary synthesis models. The
shape of the ionizing continuum is defined by the pair of values
($\nu$(Ryd), log$\nu L_{\nu}$) (Eq.~\ref{sp}).

 We have assumed the emitting gas to be located at a distance $\rm R = 500$ 
pc at the beginning of the evolution. This size is characteristic
of H{\sc ii} galaxies \citep{tel97}. This radius is subsequently
adjusted as to keep a constant stellar density configuration. This
results in all cases in a plane-parallel geometry. 

A given model is characterized by its ionization parameter U defined as:

\begin{equation} 
U = Q(H)/4\pi c n_{H} R^{2}
\end{equation}

\noindent where n$_e$ is the electron density, assumed constant for simplicity
and equal to 100 cm$^{-3}$, and $Q(H)$ is the number of Lyman ionizing
photons, calculated from the cluster SED, which reach the gas at a
velocity $c$. Finally, the gas elemental abundances are those reached
at the end of the starburst previous to the current one, as calculated
from the chemical evolution model, except for Na, Ar and Ni which are
not computed in the model and are scaled to the solar ratio \citep{asp05}.

\section{Results}

\subsection{Chemical evolution}


\begin{figure*}
\resizebox{0.6\hsize}{!}{\includegraphics[angle=-90]{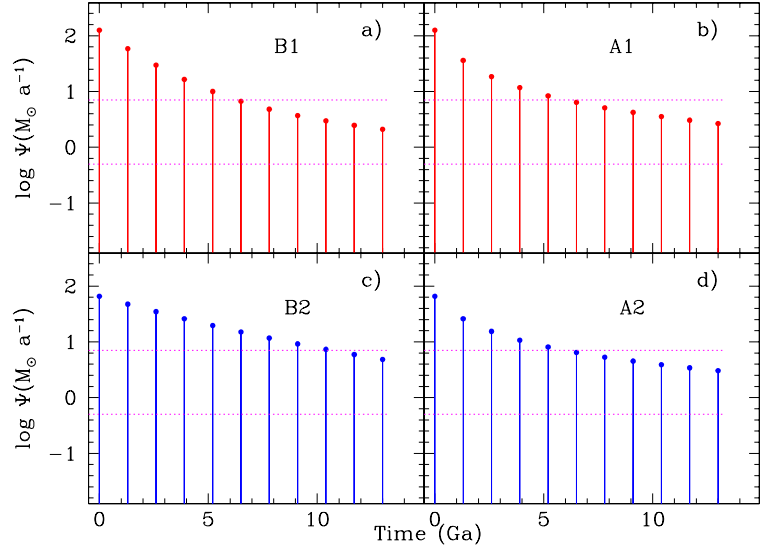}}
\caption{Star formation history for: a) burst model 1 (B1); b) attenuated
burst model 1 (A1); c) burst model 2 (B2); and d) attenuated burst
model 2 (A2). Dotted lines represent the upper and lower limits to the SFR estimated for BCG and/or H{\sc ii} galaxies by \citet{hoy04}.}
\label{sfr} 
\end{figure*}


\subsubsection{Star formation rates}

The star formation rate is one of the main results of the different
models since it drives the behaviour of all the other quantities.  In
Fig.~\ref{sfr} the star formation history for each of the 4 computed
models is shown. In all of them the first burst is strong, while the
successive bursts are less intense since the amount of gas available
has decreased in spite of the ejected gas produced by the massive
stars.  In models A (right panels), the first burst has the same star
formation rate as in Models B (left panels), since the initial
efficiency is the same for both.  The successive star formation
episodes have a smaller star formation rate in models A than the
corresponding ones in models B due to the attenuation factor included
in the inputs.

\citet{hoy04} have estimated the star formation rates for the sample of
39 local H{\sc ii} galaxies of different morphologies from
\citet{tel97}, which were selected from the most luminous of the
\citet{ter91} catalog and with H$\beta$ equivalent widths between 30
and 280 \AA\ . These star formation rates were obtained from the
H$\alpha$ luminosity, by assuming T$_{e}=10^{4}$ K and case B
recombination. They are in the range from 0.5 to 7 M$_{\odot}a^{-1}$,
marked by dotted lines in Fig.~\ref{sfr}, indicating that these H{\sc ii} 
galaxies have high star formation rates.
These findings are in agreement with those from \citet{tay94} who
found star formation rates in the range of 10$^{-4}$ to 10
M$_{\odot}a ^{-1}$.  
Our models produce SFR between the observed values after the sixth burst, 
which corresponds to 6.5 Ga after the beginning of star formation, except model B2, 
where every burst before the ninth-tenth show higher SFR than observed.


%
%

\begin{figure*}
\resizebox{0.6\hsize}{!}{\includegraphics[angle=-90]{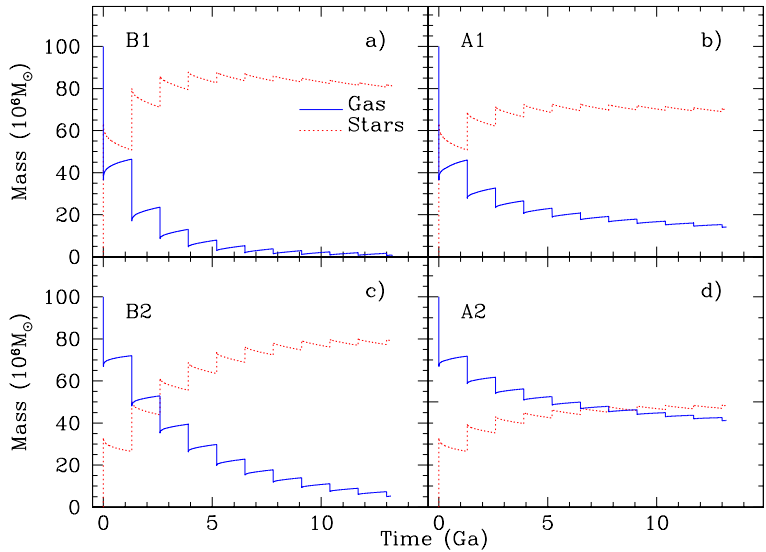}}
\caption{Time evolution of the gas and star mass for models: a) B1; b) A1; c)
B2; and d) A2.} \label{mass} 
\end{figure*}


Since the star formation consumes gas, the averaged gas mass decreases
with time following Eq.(6), as can be seen in Fig.~\ref{mass}. The
chemical evolution code, however, takes into account the mass of gas
ejected from massive stars during their evolution and, therefore, the
gas mass increases during a given burst. On the other hand, the mass
of stars which increases in each burst and globally due to the star
formation, decreases during each burst as the most massive stars end
their lives.  It is interesting to note that models with smaller
efficiencies, H, in the lower panels of Figs.~\ref{sfr} and \ref{mass}, 
consume less gas in each burst and, as a consequence, the
star formation rate remains higher, and declines more slowly, than in 
models with higher values of H shown in the upper panels.

\subsubsection{Elemental abundances}

\begin{figure*} 
\resizebox{0.5\hsize}{!}{\includegraphics[angle=0]{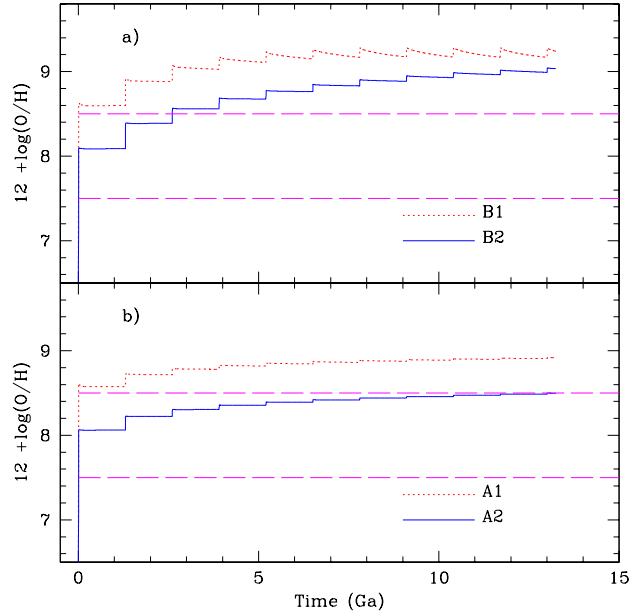}}
\caption{Time evolution of the oxygen abundance, given as
$12+log(O/H)$, for the 4 kinds of computed models: a) models of type B burst
modes) and b) models of type A attenuated models) with dotted lines
corresponding to models of type 1 and solid lines corresponding to
models of type 2.  The dashed lines limit the range of data taken from
\citet{ter91} and \citet{hoy06}.}
\label{oh}
\end{figure*}

The element content of the gas is usually measured through the oxygen
abundance, given as $\ 12+log(O/H)$. Oxygen contributes more than
 40 per cent to the mass in metals and is mainly created within massive
stars. Since these stars are very short lived, their products are ejected very
rapidly to the interstellar medium (ISM) and hence the value of
its oxygen abundance increases with time, as shown in Fig.~\ref{oh}.
In that figure we see how O/H increases abruptly when a burst takes places
remaining more or less constant between each two of them.

Observational data from \citet{ter91} and \citet{hoy06} show that most
of the H{\sc ii} galaxies have metallicity distributions between $\rm
7.5 < 12+log(O/H) < 8.5$. These limits are shown as dashed lines in
Fig.~\ref{oh}.  The first limit is reached with the first burst in all
cases. It can be seen that models of type 1 (dotted lines)
show oxygen abundances higher than the observed upper limit. The
efficiency of the first burst is so high that a large number of
massive stars are formed, and hence the oxygen abundance reaches
rather high values in a very short time. Also the oxygen abundance of
model B2 reaches values higher than shown by data from the third burst
onwards ($12+log(O/H) > 8.5$). Only model A2 shows oxygen abundances
within the range of H{\sc ii} galaxies during the whole evolution
since the attenuation of the bursts keeps the star formation lower
than in the other three cases.
The star formation efficiency taken
for model A2 results to be an upper limit for HII galaxies in oxygen
abundances and it may reproduce the most metal rich systems.

%

\begin{figure*} 
\resizebox{0.6\hsize}{!}{\includegraphics[angle=0]{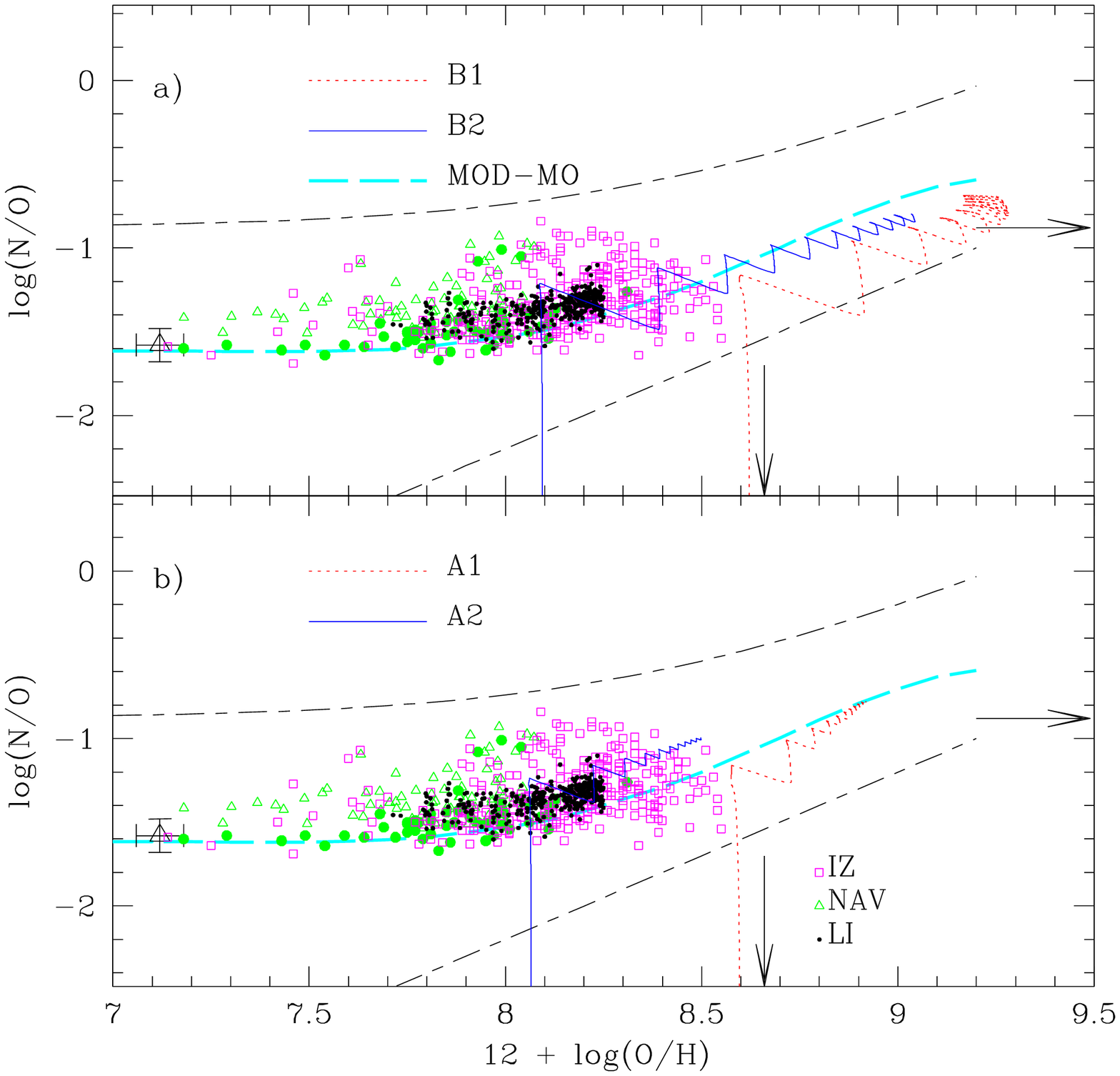}}
\caption{The logarithmic N/O ratio as a function of the oxygen
abundance, given as $12+log(O/H)$, for our 4 kinds of computed models: a) type
B models; b) type A models. The (black) dashed lines limit the range of
data shown by extragalactic H{\sc ii} regions. The (green) dots and
(magenta) squares correspond to the derived values for low metallicity
galaxies from \citet{izo99,izo05,izo06,hoy06,lia06,nava06}, while the
large triangle is the value for the lowest metallicity galaxy in the
sample {SBS 0335-052W}. The long dashed (cyan) lines represents the
averaged trend shown by chemical evolution models from \citet{mol06}
for spiral and irregular galaxies with continuous star formation. The
arrows mark the solar values.} 
\label{no} 
\end{figure*}

Abundance ratios also give information about the star formation
process, essentially the time scale or duration of the bursts, when
the production of the elements involved takes place in stars in
different mass ranges. This is the case for the N/O ratio. Oxygen is
very quickly ejected by the most massive stars. Nitrogen, however, may
be created in stars of all masses. Moreover, the production of
nitrogen may take place in two ways: 1) the main process requires
another element like carbon or oxygen, to synthesize N through the CNO
cycle, thus N shows, at least in part, a secondary behavior. In that
case the N abundance has to be proportional to the initial abundance
of heavy elements and hence the N/O ratio grows with metallicity.
There is, however, a proportion of Nitrogen which must have a primary
origin, as suggested by the data in Fig.~\ref{no}
\citep{izo99,izo05,izo06} shown by (green) dots which exhibit a
constant N/O ratio. Nitrogen may be produced by low and intermediate
mass stars \citep{gav06} and therefore its contribution may appear in
the ISM a certain time after the massive stars have died.  In fact, N grows
more slowly than the oxygen abundance in the range $7.6 < 12+log(O/H)< 8.2$, 
which may be explained by this primary behavior \citep[see][ for
details]{mol06}. In Fig.~\ref{no} we can see the oscillating behavior
of the N/O ratio shown by our computed models due to the successive
bursts of star formation followed by quiescent periods.  At the
beginning of a given burst, the N/O ratio decreases, as the O is
created and ejected by the most massive stars in the burst; then it
increases during the quiescent periods between bursts as the low and
intermediate mass stars die and eject their synthesized nitrogen while
the oxygen abundance remains constant.

According to the results of this section, model A2 is the one better
reproducing the observations, since the total mass, the gas fraction,
the star formation rate and the oxygen abundances are within the range
shown by data. Therefore, from now onwards, in what regards the
ionizing continuum and emission line properties only the A2 model will
be analyzed.


\subsection{Evolutionary stellar population synthesis}

The convolution of the results of the chemical evolutionary code with
the single stellar populations of the synthesis evolutionary code
gives us directly the SED, $\rm L_{\lambda}$, in time steps of logt$=$ 0.05.

From this SED the number of hydrogen ionizing photons, coming from hot
and young stars of the first spectral types (O-B), can be calculated
as:
\begin{equation} 
Q(H)= \int_{\nu_{0}}^{\infty} \frac{L_{\nu}}{h\nu}d\nu
\end{equation}
where L$_\nu$ is the continuum luminosity at frequency $\nu$.

\begin{figure*} 
\resizebox{0.5\hsize}{!}{\includegraphics[clip,angle=0]{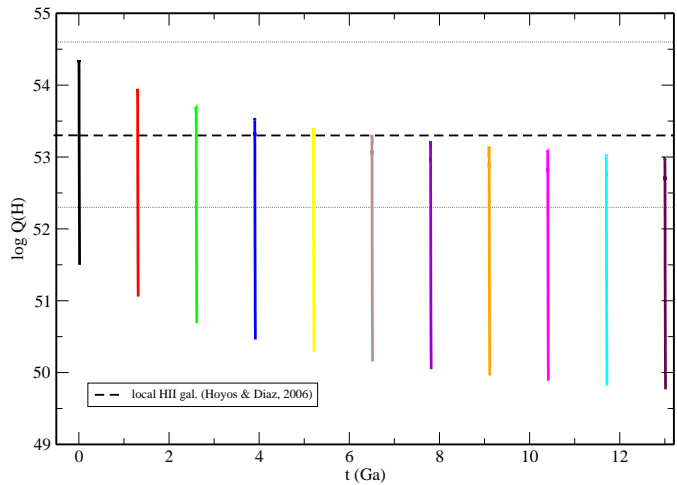}}
\caption{The time evolution of the hydrogen ionizing photons, Q(H),
for model A2. The dotted line marks the mode of the distribution for
local H{\sc ii} galaxies found by \citet{hoy06}. Each coloured line
indicates the evolution within a burst. Different colour lines are used
for different bursts.}
\label{foton} 
\end{figure*}

Fig.~\ref{foton} shows the number of hydrogen ionizing photons as a
function of time for model A2.  In the first burst, with a total mass
of $\sim 30 \cdot 10^{6}$ M$_{\odot}$ forming stars, the number of
massive ionizing stars is high and so it is the number of ionizing
photons.  The number of ionizing photons per unit stellar mass in our
models is ${\log(Q(H)/M_{\odot})} \sim 46.8$ at the starting of the first
burst, corresponding to a Zero Age Stellar Population (ZASP) of very
low metallicity and then decreases as the cluster ages. This initial
number of ionizing photons per unit stellar mass is almost the same
for the successive bursts although, decreasing slightly as the abundance
increases.

\begin{figure*} 
\resizebox{0.5\hsize}{!}{\includegraphics[clip,angle=0]{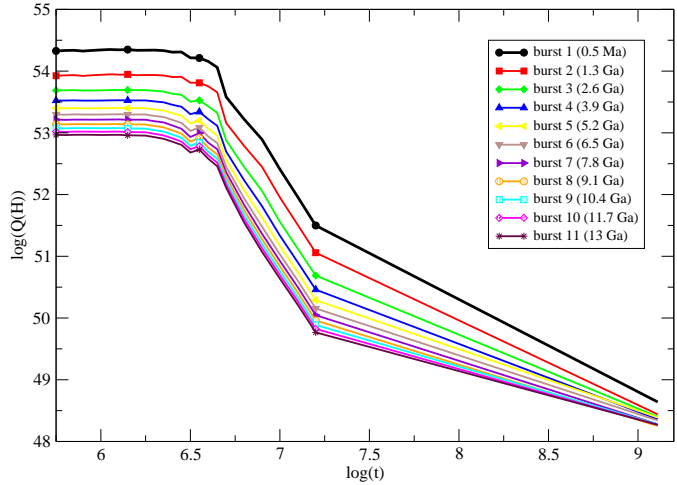}}
\caption{The time evolution of the number of ionizing photons for each
of the successive bursts during the 1.3 Ga inter-burst period. The
burst number increases downwards.  Different colour lines are used for
the different bursts.}
\label{foton-burst} 
\end{figure*}

Within each burst the number of ionizing photons decreases by almost
two orders of magnitude from 0.5 to 10 Ma, as show in
Fig.~\ref{foton-burst}, and by almost 8 orders of magnitude by 100 Ma.
Essentially no ionizing photons are available during the quiescent
inter-burst periods. At the beginning of each successive burst, the
number of ionizing photons raises abruptly but reaching a value
progressively lower as less gas is available to form stars. Also, as
the mean gas metallicity increases with time, the number of ionizing
photons per unit stellar mass decreases \citep{garc95a,garc95b}. This effect
increases with the age of the ionizing cluster reaching a factor of
about 5 at 10 Ma.


The average number of ionizing photons observed for local H{\sc ii} galaxies
is $\log{(Q(H))}= 53.3$ \citep{hoy06}. This number of ionizing
photons, marked as a dotted line in Fig.~\ref{foton}, is reached by
model A2 at around the fifth or sixth burst of star formation, that is
for the burst occurring 5.2-6.5 Ga after the beginning of the formation of
the galaxy, when the predicted star formation rate also fits the
data (see Fig.~\ref{sfr}). On the other hand, the average derived
ionization parameter for the local H{\sc ii} galaxy sample is $log U =
-2.5$ which in our models corresponds to an age of the ionizing
population of about 6.5 Ma. It should be noted that during the first 7
Ma of a burst the number of ionizing photons remain essentially
constant (see Fig.~\ref{foton-burst}).

Therefore, if the current burst is identified with redshift $z=0$, the
formation of the galaxy would have taken place at a redshift $z\sim 0.7$ 
\citep[ using $H_{0}=74$ km.s$^{1}$.Mpc$^{-1}$, ][]
{macri06}. Also the observed number of ionizing photons for LCBGs at a
redshift $0.46<z<0.7$, estimated as $\log{(Q(H))}\sim 54.6$
\citep{ham01}, would correspond to the value of the first burst of our
models.  According to these results, our model seems consistent with
observations at these intermediate redshifts reproducing at the same
time those corresponding to the more luminous local H{\sc ii}
galaxies.

The question of if these galaxies are young galaxies, experiencing
their first burst of star formation or if, on the contrary, there
exists an underlying population formed some Gas ago, may be addressed
by multi-colour photometry. Although a direct detection of the
underlying population in LCBG or H{\sc ii} galaxies is complicated
because the observed light is dominated by massive stars, the effect
of an underlying stellar population should be easily seen in the
observed colours of these galaxies, since any star forming burst
previous to the currently observed one will contribute substantially
to the total continuum luminosity at the different wavebands.


\begin{figure*}
\resizebox{0.9\hsize}{!}{\includegraphics[clip,angle=0]{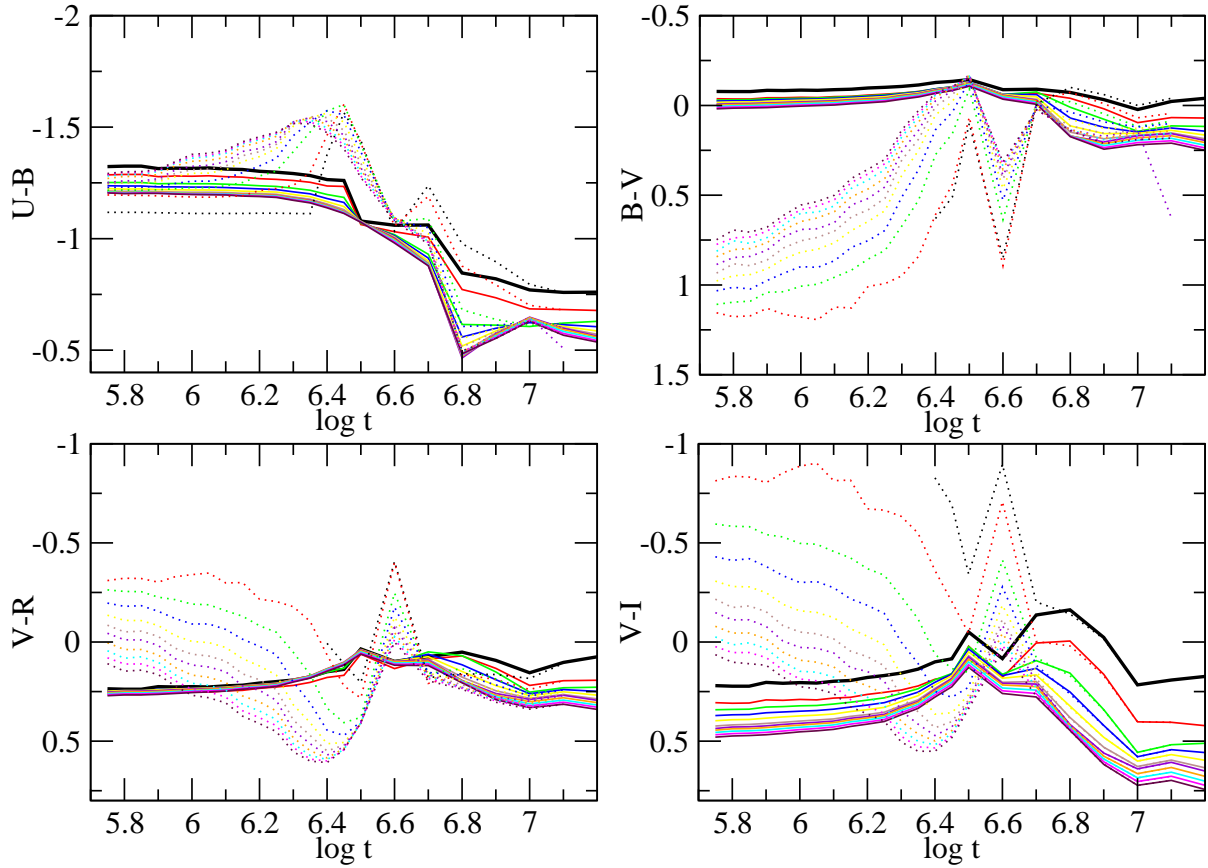}}
\caption{Time evolution of colours for model A2: a) U-B; b) B-V; c)
V-R; d) V-I. The meaning of the (coloured) lines for different bursts
is the same as in Fig.~\ref{foton}. Solid lines show the pure
continuum colours while dotted lines also include the contribution by
the strong emission lines.}
\label{colors} 
\end{figure*}


In Fig.~\ref{colors} we represent the time evolution of the continuum U-B, B-V,
V-R and V-I colours for our model A2 along the first 10 Ma after each
burst.  The first burst --solid black thick line-- follows the
expected evolution of a single stellar population of very low
metallicity, remaining bluer than during the subsequent bursts where the
effect of the accumulated continua from the previous star formation
episodes makes colours become redder than expected for a SSP. Even
though they are also starbursts in themselves, their star formation
efficiency is much lower than that of the first burst, and so is their
contribution to the total continuum luminosity which results in
redder colours.  Furthermore, the metallicity, almost zero for the
first burst, increases up to $\rm Z=Z_{\odot}/5$ already during the second
burst which also has an reddening effect over the colours.

\subsection{Photo-ionization models}

The calculated SED in each time step can be used as input ionizing
sources in the photo-ionization code to predict the emission lines
intensities which provide information about the youngest stellar
population which dominates the final spectra.

\begin{figure*} 
\resizebox{0.6\hsize}{!}{\includegraphics[clip,angle=0]{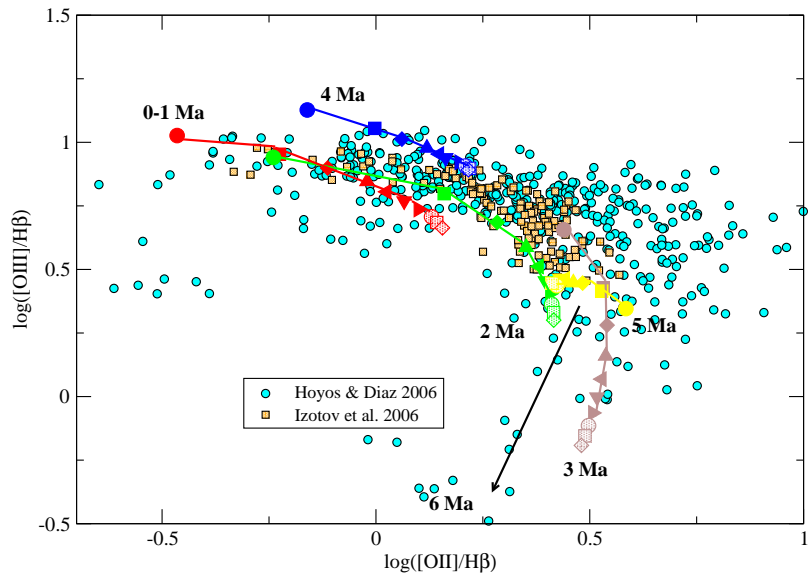}}
\caption{The relation between the oxygen line ratios for model A2.
Solid coloured lines join the values reached at the same age by each
different burst.  Red, green, brown, blue and yellow colours
correspond to ages 0-1, 2, 3, 4 and 5 Ma respectively.Different
bursts, except the first one, are represented by different symbols:
solid circles, squares, diamonds, triangles up, triangles left,
triangles down and triangles right correspond to bursts at 1.3, 2.6,
3.9, 5.2, 6.5, 7.8 and 9.1 Ga respectively. Open circles, squares and
diamonds correspond to bursts at 10.4, 11.7 and 13.2 Ga
respectively. The data are from references cited in the text.}
\label{o3o2}
\end{figure*}

Fig.~\ref{o3o2} shows the relation between the $[OIII]\lambda\lambda$
4959,5007/H$\beta$ and [OII]$\lambda \lambda$ 3727,3729/H$\beta$ ratios,
which constitutes an excitation diagnostics for ionized nebulae.  In
this diagram, the most excited objects lie up and to the left while
the objects with the lowest excitation are at the bottom right.

The data shown in the graph have been extracted mainly from two main sources. 
First, the compilation from \citet{hoy06} which provides emission
line measurements, corrected for extinction, published for local 
H{\sc ii} galaxies.  The sample comprises 450 objects and constitutes a
large sample of local H{\sc ii} galaxies with good-quality spectroscopic
data. The sample is rather inhomogeneous in nature, since the data
proceed from different instrumental setups, observing conditions and
reduction procedures, but have been analysed in a uniform way. Data
for these sample objects include the emission line intensities of:
[OII]$\lambda\lambda$ 3727,29 \AA\ , [OIII]$\lambda\lambda$ 4959,5007
\AA\ , and [NII]$\lambda\lambda$ 6548,84 \AA\ , all of them relative to
H$\beta$, and the equivalent widths of the [OII] and [OIII] emission
lines -- EW([OII]) EW([OIII]) --, and the H$\beta$ line, EW(H$\beta$).
As a second source, we have used the metal poor galaxy data from the
Data Release 3 of Sloan Digital Sky Survey, taken from
\citet{izo06}. The Sloan Digital Sky Survey \cite[][]{yor00}
constitutes a large data base of galaxies with well defined selection
criteria and observed in a homogeneous way.  The SDSS DR3 \citet{aba05}
provides spectra in the wavelength range from 3800 to 9300 \AA\ for
$\sim$ 530000 galaxies, quasars and stars.  \citet{izo06} extracted
$\sim$ 2700 spectra of non-active galaxies with the
[OIII]$\lambda$4363 \AA\ emission detected above 1$\sigma$ level. This
initial sample was further restricted to the objects with an observed
flux in the H$\beta$ emission line larger than $10^{-14} erg s^{-1} cm^{-2}$ 
and for which accurate abundances could be derived. They have
also excluded all galaxies with both [OIII]$\lambda$4959/H$\beta$ $<$ 0.7 
and [OII]$\lambda$3727/H$\beta$ $>$ 1.0.  Applying all these
selection criteria, they obtain a sample of $\sim$ 310 SDSS
objects. Data for these sample objects include the emission line
intensities of: [OIII]$\lambda$4959,5007 \AA\ and [NII]$\lambda$6584
relative to H$\beta$ and the equivalent width of H$\beta$.  They also
include the intensity of the [OII] $\lambda\lambda$ 3727,29 \AA\
emission line for the lowest redshift objects.

The very young stellar population during the first Ma after each burst
produces a high ionization, so the [OIII]/H$\beta$ ratio is high, then
it decreases raising again at 4 Ma due to the presence of Wolf Rayet
stars that produces a harder continuum \citep{garc95a,gmb98}. After 5
Ma, the burst evolves to lower values of $log[OIII]/H\beta$. Diagonal
lines in this plot correspond to nearly constant ionization parameter
and are swept along by the different successive bursts of a given age.
High excitation objects, with high $log[OIII]/H\beta$ and low
$log[OII]/H\beta$ ratios, have low metallicities and they are not
reproduced by our model, which fits better the high metallicity
sample, in the middle to right side of the panel.

%
%

\begin{figure*} 
\resizebox{0.6\hsize}{!}{\includegraphics[clip,angle=0]{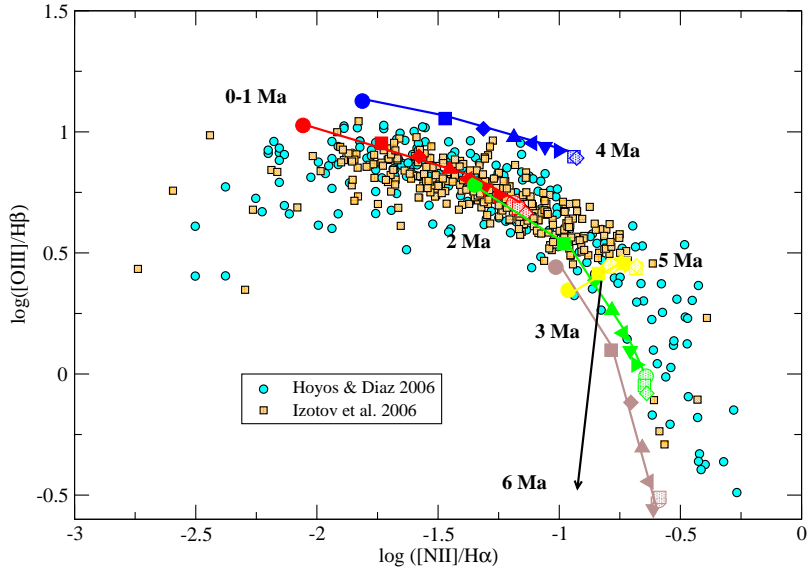}}
\caption{The relation between the excitation parameter, 
$log([OIII]/H\beta)$, and the metallicity indicator $log([NII]/H\alpha)$
for model A2. The meaning of colours, symbols and lines is the same as
in the previous figure.}
\label{nho3} 
\end{figure*}


It can be better seen in Fig. ~\ref{nho3}, which shows the relation between the excitation parameter, $log([OIII]/H\beta)$, and the metallicity indicator
$log([NII]/H\alpha)$ which gives information about the younger
populations plus the star formation rate \citep{ken94}.
 At the beginning of the evolution, as we saw in previous sections, the star
formation rate is high, and the H$\alpha$ emission of our model is
strong. As the galaxy evolves, the H$\alpha$ emission decreases while
the [NII] emission increases due to the growth of metallicity.

Our models reproduce the general trend shown by data but cannot
reproduce the lower metallicity (low $[NII]/H_{\alpha}$ ratio) objects
to the left of the graph.  This is due to the fact that the efficiency
of the first burst, 33.1 per cent, produces a metallicity which is already
too high. A lower efficiency is therefore required to reproduce the
observations of the less metallic objects.

\subsection{Combined continuum and emission line colours}

Once the continuum spectral energy distributions and the different 
emission line fluxes are computed, we can calculate the contribution 
of these emission lines to the different broad band filters and then 
synthesize the colours of our model galaxies. These are the colours 
which are readily observable through integrated photometry.

We have done so taking into account only the strongest emission lines
that contribute to the colour in each broad-band spectral interval at
redshift zero. These are: [OII] $\lambda\lambda$ 3727, 3729 \AA\ in U,
H$\beta$ $\lambda$ 4861 \AA\ in B, [OIII]$\lambda\lambda$ 4959, 5007
\AA\ in V, H$\alpha$ $\lambda$ 6563 \AA\ in R and [SIII]
$\lambda\lambda$ 9069, 9532 \AA\ in I.

The continuum colours and the colours including the emission line contribution 
at redshift zero are given in columns 3 to 10 of Table 1 for each 
of the models characterized by their burst number and age (columns 1 and 2).
Columns 11 to 15 of the same table give the contribution, in percentage, by 
the emission lines to the different continuum fluxes. 

The necessary information to calculate these contributions for any
other redshifts in the nearby universe (z $ \lesssim $0.1), is given
in Table 2 which lists for each of the models characterized by their
burst number and age (columns 1 and 2), the continuum fluxes in each
of the U,B,V,R and I (columns 3 to 7) and the fluxes in the [OII],
H$\beta$, [OIII], H$\alpha$ and [SIII] lines (columns 8 to 12).

\begin{figure*} 
\resizebox{0.65\hsize}{!}{\includegraphics[clip,angle=0]{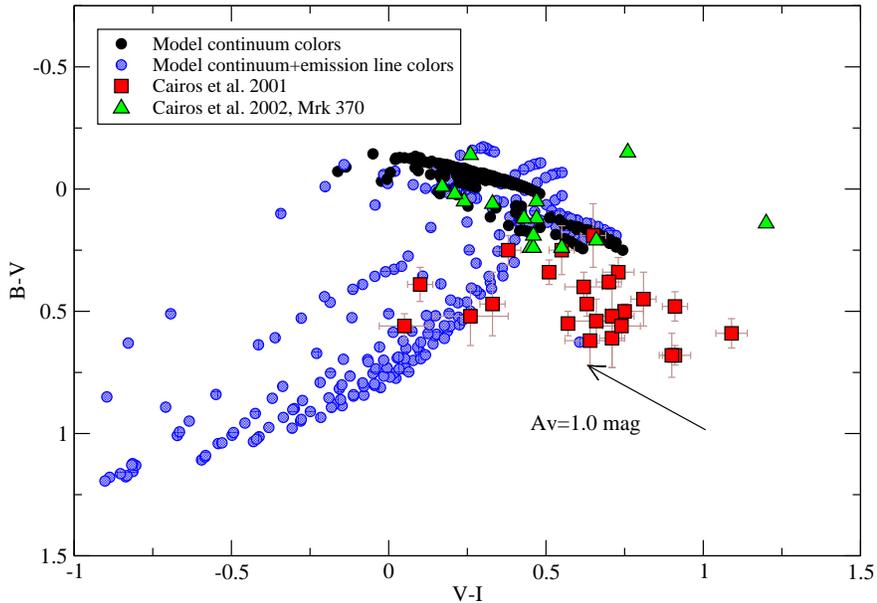}}
\caption{B-V vs V-I colour-colour diagram for model A2. Light blue
circles show the continuum colours (solid lines in Figure~\ref{colors}) 
while black solid ones show the colours computed including the
contribution by the stronger emission lines. The data by \citet{cai011}
on BCG and \citet{cai02} in Mrk~370 are included as red squares and
green triangles respectively.}
\label{colours-lines}
\end{figure*}

Figure ~\ref{colors} shows, with dotted lines, the new computed
colours including the contribution by the different emission lines. As
it can be seen the colours do not change by a large amount in the case
of U-B or V-R. However the effect in the B-V and V-I colours is rather
dramatic. Figure \ref{colours-lines} shows the B-V vs V-I
colour-colour diagram for model A2. Light blue circles show the
continuum colours (solid lines in Figure \ref{colors}) while black
solid ones show the colours computed including the contribution by the
stronger emission lines given in Table 1. The inclusion of the
contribution by the emission lines to the continuum colours shifts the
position of the model points almost perpendicularly to the originally
computed ones. The location of blue points is mainly
determined, even at zero redshift, by the contribution of strong
[OIII]4959,5007 emission lines to the V band.  In the same diagram we
show the data by \citet{cai011,cai012} on BCG as red squares with
error bars. These correspond to readily observed colours therefore
including both continuum and line emission. No reddening correction
has been applied. It can be seen that some of the data are impossible
to reproduce by the models which do not take into account the
contribution by emission lines, even if some amount of reddening is
invoked (the arrow in the diagram shows the correction corresponding
to 1 mg visual extinction), while a certain amount of reddening would
bring most of the data in agreement with the models. On he other hand
the data by \citet{cai02} of resolved locations in Mrk~370, shown in
the figure as green triangles, correspond to data that have been
corrected for the contribution by emission lines and therefore
correspond to pure continuum colours. In these case, the agreement
with the models is excellent.


\section{Discussion}

The evolution of H{\sc ii} galaxies must be considered by looking
simultaneously at both continuum and emission line properties, the
first one providing information on a long time scale evolution, of the
order of Ga, and the second one being related to the evolution in a
much shorter time scale, of the order of Ma.

\begin{figure} 
\resizebox{\hsize}{!}{\includegraphics[clip,angle=0]{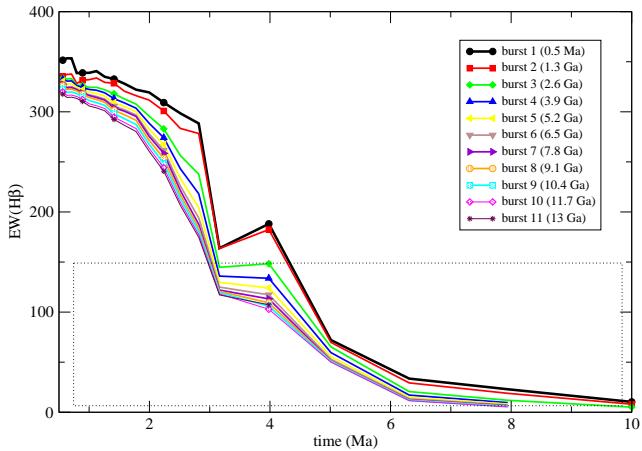}}
\caption{Time evolution of the H$\beta$ equivalent width of model A2. The
thick line corresponds to the first star burst of our model, a SSP
with very low metallicity. Thin lines represent the successive bursts with
same colours as in Fig.~\ref{foton}.} 
\label{hbeta}
 \end{figure}

 \citet{dot81} suggested the use of the equivalent width of H$\beta$,
EW(H$\beta$), as an age estimator for H{\sc ii} regions, and applied
this method to rank the ages of H{\sc ii} regions in the Magellanic
Clouds. If a SSP is considered, the H$\beta$ emission is very high at
the beginning of the burst, while the continuum at H$\beta$, dominated
by the most massive and luminous stars, is low hence producing a large
value of EW(H$\beta$). As the burst evolves, the H$\beta$ luminosity
decreases as does the number of ionizing photons, and the contribution
to the continuum increases thus lowering the value of
EW(H$\beta$). This case corresponds to the thick (black) line in
Fig.~\ref{hbeta}.


The evolution of the equivalent width in our model is shown in
Fig.~\ref{hbeta}. It is seen that the first burst, that corresponds to
a SSP, has an initial equivalent width slightly greater than the
successive bursts, which decreases with time reaching very low values
($<$ 40 \AA\ ) 6 Ma after the beginning. When a new burst takes place,
new massive ionizing stars form and the EW(H$\beta$) goes up again,
although not as much as in the previous one due to the decreasing gas
mass involved in the burst, increased metallicity and the accumulated
contribution from the continua from the previous bursts. At any rate,
even taking into account this contribution, the EW(H$\beta$) remains a
good age indicator for the age of the current burst, inside 10 Ma. On
the other hand it is not possible to detect the presence of the
underlying stellar populations for a given galaxy solely on the basis
of this parameter.

The distribution of the EW(H$\beta$) in H{\sc ii}
galaxies shows that most of them have values lower than 150 \AA\
\citep{ter91,hoy06} which, in our models, corresponds to ages greater
than 3-4 Ma. \citet{ter04}, from the application of inversion methods,
have shown that, globally, this distribution is inconsistent with what
is expected from a single burst scenario and resembles more the
distribution obtained for a succession of short bursts separated by
quiescent periods with little or no star formation.


\begin{figure}
\resizebox{\hsize}{!}{\includegraphics[clip,angle=0]{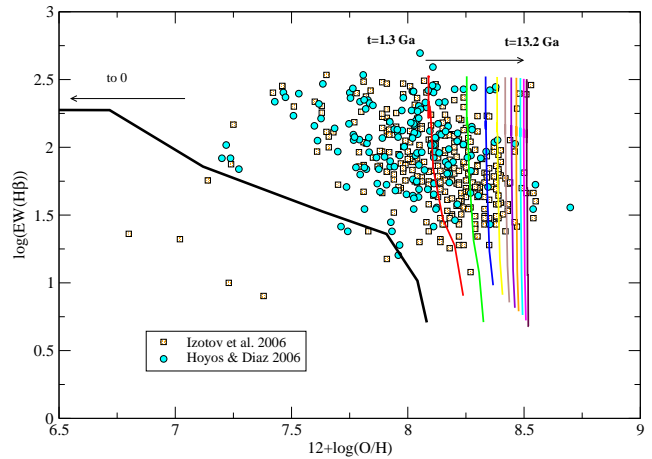}}
\caption{EW(H$\beta$) as a function of the metallicity $12+log(O/H)$
for the model A2. Colours have the same meaning than in Fig.~\ref{colors}.
The first burst (solid black line) does not appear in the graph because
its abundance is still the primordial one, $12+log(O/H) << 7$. }
\label{hbeta_ox} 
\end{figure}





These results are not that strong when samples showing a restricted
metallicity range are analyzed. However, as we have seen before,
metallicity increases with time in a way which constrains the SFH which
is best reproduced by our A2 model.  Fig.~\ref{hbeta_ox} shows the
evolution of the $\rm EW(H\beta)$ with oxygen abundance.  During the
duration of the first burst the $\rm EW(H\beta)$ decreases while the
oxygen abundance increases from zero to $ 12 + log (O/H) \sim 8$. The
successive bursts start with this higher abundance and each of them
evolves vertically in a 10 Ma time scale with the $\rm EW(H\beta)$
decreasing rapidly at almost constant oxygen abundance.  This occurs
because the oxygen is produced by the most massive stars in a very
short time and therefore its abundance changes very little during the
next 10 Ma after each burst (see Fig.~\ref{oh}).  It is clear that our
model is able to reproduce the most metallic galaxies in the samples
taken from \citet{hoy06} and \citet{izo06} where this high abundance is
explained as the consequence of the gas being processed by one or more
previous generations of stars.  An alternative scenario to produce
such high oxygen abundances would be the occurrence of a very intense
initial star formation burst, with a SFR of the order of 100 M$_{\odot}$
Ma$^{-1}$, out of the range observed in local H{\sc ii} galaxies (see
Figures \ref{oh} and \ref{sfr}).
On the other hand, galaxies $12+log(O/H) \le 8$ are not reproduced by
our model.  In order to fit these data, a model with a star
formation efficiency lower than assumed, that is, $ <$ 33
per cent, should be used.

%
 
\begin{figure*}
\resizebox{0.6\hsize}{!}{\includegraphics[clip,angle=0]{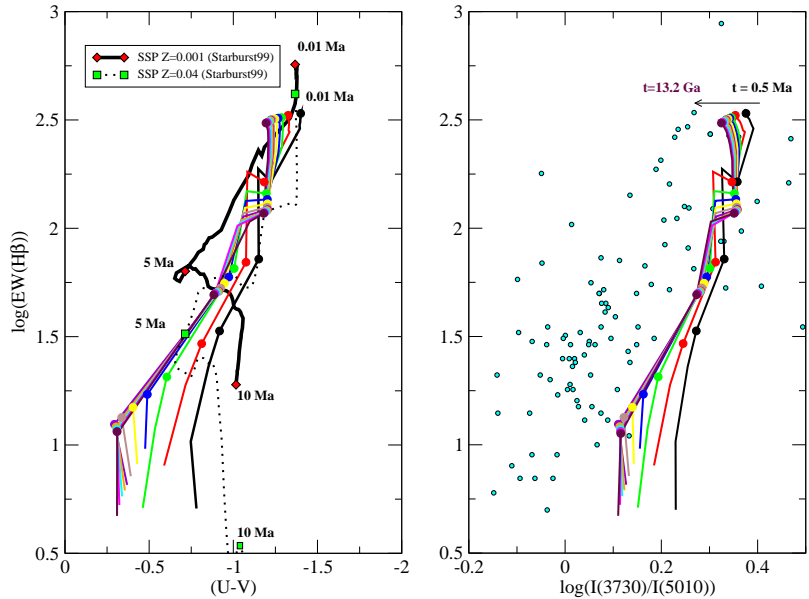}}
\caption{Left panel: The equivalent width of H$\beta$ as a function of
the continuum colour for model A2. The results for the SSPs from the
model Starburst99 \citet{stb99} with two different metallicities as
labelled are overplotted. Colours have the same meaning as in
Fig.~\ref{colors}. Right panel: The equivalent width of H$\beta$ as a
function of the pseudo-colour defined by equations 13 to 15 for model
A2.  The data are from \citet{ter91} and \citet{sal95}. In both panels
the large dots mark ages separated 2 Ma.}
\label{beta-cont} 
\end{figure*}

The most informative data to uncover the presence of underlying
populations in H{\sc ii} galaxies consist of the combination of a line
emission parameter, which characterizes the properties of the current
burst of star formation, and a continuum colour which represents
better the SFH in a longer time scale.  Fig.~\ref{beta-cont}, left
panel, shows the equivalent width of H$\beta$ {\sl vs} the U-V colour
along the evolution of the galaxy for the successive stellar
bursts. When a given burst occurs the EW(H$\beta$) is high and
decreases as the stellar population ages.  We have over-plotted the
results for the SSPs from the model Starburst99 \cite[][SB99]{stb99}
with two different metallicities as labelled.  As expected, our
results agree better with the higher metallicity SB99 model. 
The black line in the left panel, that represents a metal-poor SSP, is
too blue compared with the data for a given EW(H$\beta$).
In order to decrease EW(H$\beta$) and move the model colour to the
red, a more metal-rich SSP was selected (dotted line at the left
panel), but even such unrealistic high abundance does not reproduce the
observations, as we have shown in Fig.~\ref{oh}. Therefore, the observed
trend can not be explained as an age effect since EW(H$\beta$) for low metallicity SSP does
not reach values lower than 100\AA\ and does not reproduces the colors,
neither  as a metallicity effect because such high abundances are not
observed in H{\sc ii} galaxies.  It is therefore necessary to include
both effects simultaneously, as we have done, to see how the effect of an
underlying older population  contributing to the color of the
continuum makes it redder.
From the observational point of view, U-V colours for H{\sc ii} galaxies are
scarce, therefore we have compared our results with those from
\citet{hoy06} that include galaxies taken from \citet{ter91} and
\citet{sal95} providing the equivalent widths of both the [OII] and
[OIII] lines. We have then computed pseudo-colours from the
intensities of the adjacent continua of [OII]$\lambda$3727 and
[OIII]$\lambda$5007 lines as:

\begin{equation} 
EW_{[OII]\lambda 3727}(\rm \AA)= \frac{[OII]\lambda 3727 (erg s^{-1})}{I_{cont}(3730)(erg s^{-1}{\AA}^{-1})}
\end{equation} 
and

\begin{equation} 
EW_{[OIII]\lambda 5007}(\rm \AA)= \frac{[OIII]\lambda 5007(erg s^{-1})}{I_{cont}(5010)(erg s^{-1}{\AA}^{-1})} 
\end{equation}

\begin{equation} 
log\left( \frac{I_{cont}(3730)}{I_{cont}(5010)} \right)= log\left(\frac{[OII]\lambda 3727} {[OIII]\lambda 5007} \frac{EW_{[OIII]\lambda 5007}}{EW_{[OII]\lambda 3727}} \right) 
\end{equation}


The right panel of Fig.~\ref{beta-cont} shows the equivalent width of
H$\beta$ as a function of the pseudo-colour defined above for model A2
together with the observational data.  
It should be noted that no correction for extinction associated with the continuum light has been made. This
correction amounts to about 0.15 dex for a standard reddening law and
therefore a SSP would always be unable to reproduce the observed
colours. Yet our Model A2 produce UV continuum colours which are too blue as
compared with the bulk of the observations which indicates that our
assumed degree of attenuation is too low.

\section{Summary an conclusions}

We have explored the viability of a kind of a theoretical model in
order to understand how the star formation takes place in H{\sc ii}
galaxies.  Our work involves the combination of three different tools:
a chemical evolution code, an evolutionary synthesis code and a
photo-ionization code, in a self-consistent way, {\it i. e.} taking
the same assumptions about stellar evolution and nucleosynthesis, and
takes into account the resulting metallicity in every time step. The
chemical evolution code provides the star formation history of the
galaxies as well as the time evolution of the abundances in their
interstellar medium, that can be confronted with observations.  The
evolutionary synthesis code provides the spectrophotometric evolution
of the integrated stellar population.  Finally, the photo-ionization
code uses the properties of the ionizing stellar population to
calculated the emission line spectra of the ionized gas.

Each galaxy is modelled assuming an initial amount of unprocessed gas
of 10$^{8}$ M$_{\odot}$ in a 1 kpc diameter. The evolution is computed
along a total duration of 13.2 Ga during which 11 successive
starbursts, separated by 1.3 Ga, take place . Different types of
models have been computed with different starbursting properties:
equal and attenuated bursts, and different values of the initial star
formation efficiency. The only model which has proved to be able to
fit the observational data is model A2, which consists of attenuated
bursts with an initial star formation efficiency of 33.1 per cent.  This type
of model reproduces the star formation rate estimates of local H{\sc ii} 
galaxies and the oxygen abundances of the average metallicity
objects although fails to account for the most metal deficient ones
which would require models with lower star formation efficiencies than
assumed.

We have computed the time evolution of the continuum colours with and
without the inclusion of the emission lines. It is shown that both
nebular continuum and line emission must be taken into account
in the photometric studies of the underlying stellar populations of H{\sc ii}
galaxies. The effect of the emission lines
are more pronounced in the B-V and V-I colours.

The combination of parameters which characterize the current star
formation on a time scale of Ma, such as the equivalent width of
H$\beta$, and the SFH over a time scale of Ga, such as broad-band
continuum colours, is found to provide an effective means to uncover
the presence of underlying stellar populations.  The comparison of
models and observations show that in most H{\sc ii} galaxies SSP are unable
to reproduce the relatively red colours shown by data with the
observed H$\beta$ equivalent width values requiring the contribution of
previous stellar generations. Our model, however, produces U-V colours
that are too blue when compared with observations, therefore implying
that the successive bursts that take place in H{\sc ii} galaxies should be
even more attenuated than has been assumed in our work.
  
\section*{Acknowledgments} 
This work has been partially supported by DGICYT grant
AYA-2004-02860-C03.  AID acknowledges support from the Spanish MEC
through a sabbatical grant PR2006-0049. Also, partial support from the
Comunidad de Madrid under grant S-0505/ESP/000237 (ASTROCAM) is
acknowledged.  Support from the Mexican Research Council (CONACYT)
through grant 19847-F is acknowledged by RT. 
We thank the referee for all the comments and suggestions for a better understanding of this work.
 We thank Elena Terlevich for a careful reading of this manuscript and the hospitality
of the Institute of Astronomy in Cambridge where this paper was
partially written.

 \onecolumn
 \landscape

\begin{table*}
\centering
 \begin{minipage}{148mm}
\caption{Evolution of continuum and total broad band colors for different 
bursts along the time after each stellar burst occurs.
Only the last burst evolution is shown. The whole table is
available in electronic format.}
 \begin{tabular}{ccccccccccccccc}
\hline
NB & log$\frac{t}{a}$ & (U-B)$_c$ & (B-V)$_c$ & (V-R)$_c$ & (V-I)$_c$
& (U-B) & (B-V) & (V-R) & (V-I) & \% U & \% B & \% V & \% R & \% I \\
\hline
11  &  5.75  & -1.205  &  0.019  &  0.271  &  0.480  & -1.292  &  0.748  & 0.109   &  0.064   & 21.254  & 14.681  & 57.230   & 50.400   & 37.264   \\       
11  &  5.80  & -1.202  &  0.015  &  0.267  &  0.474  & -1.300  &  0.694  & 0.147   &  0.102   & 21.845  & 14.497  & 54.820   & 49.812   & 36.728   \\       
11  &  5.85  & -1.203  &  0.014  &  0.267  &  0.472  & -1.301  &  0.706  & 0.142   & -0.065   & 21.900  & 14.506  & 55.610   & 50.230   & 42.959   \\       
11  &  5.90  & -1.203  &  0.012  &  0.264  &  0.468  & -1.311  &  0.678  & 0.164   &  0.117   & 22.634  & 14.424  & 54.460   & 50.115   & 37.121   \\       
11  &  5.95  & -1.200  &  0.008  &  0.260  &  0.461  & -1.345  &  0.611  & 0.222   &  0.177   & 25.040  & 14.260  & 51.674   & 50.000   & 37.274   \\       
11  &  6.00  & -1.198  &  0.004  &  0.255  &  0.453  & -1.378  &  0.555  & 0.269   &  0.225   & 27.207  & 13.997  & 49.152   & 49.839   & 37.295   \\       
11  &  6.05  & -1.197  &  0.001  &  0.252  &  0.448  & -1.383  &  0.535  & 0.281   &  0.236   & 27.391  & 13.889  & 48.246   & 49.628   & 37.110   \\       
11  &  6.10  & -1.196  & -0.002  &  0.248  &  0.441  & -1.397  &  0.500  & 0.307   &  0.260   & 28.277  & 13.716  & 46.603   & 49.456   & 36.942   \\       
11  &  6.15  & -1.191  & -0.009  &  0.238  &  0.426  & -1.435  &  0.407  & 0.374   &  0.327   & 30.679  & 13.226  & 41.893   & 48.686   & 36.373   \\       
11  &  6.20  & -1.189  & -0.015  &  0.232  &  0.414  & -1.457  &  0.355  & 0.412   &  0.363   & 31.957  & 12.892  & 39.095   & 48.422   & 36.106   \\       
11  &  6.25  & -1.185  & -0.021  &  0.224  &  0.401  & -1.477  &  0.300  & 0.452   &  0.399   & 33.171  & 12.523  & 36.012   & 48.132   & 35.917   \\       
11  &  6.30  & -1.172  & -0.035  &  0.202  &  0.367  & -1.517  &  0.136- & 0.554   &  0.504   & 35.548  & 11.422  & 25.592   & 46.160   & 34.394   \\       
11  &  6.35  & -1.160  & -0.047  &  0.182  &  0.335  & -1.515  &  0.027- & 0.601   &  0.551   & 35.367  & 10.370  & 17.718   & 44.104   & 32.609   \\       
11  &  6.40  & -1.138  & -0.067  &  0.148  &  0.278  & -1.475  & -0.068- & 0.608   &  0.551   & 33.098  &  8.712  &  9.986   & 41.101   & 30.006   \\       
11  &  6.45  & -1.113  & -0.083  &  0.116  &  0.226  & -1.403  & -0.107- & 0.543   &  0.483   & 29.051  &  7.251  &  6.539   & 36.974   & 26.347   \\       
11  &  6.50  & -1.073  & -0.107  &  0.061  &  0.128  & -1.274  & -0.152- & 0.407   &  0.336   & 20.862  &  4.735  &  2.161   & 28.889   & 19.217   \\       
11  &  6.60  & -1.020  & -0.076  &  0.084  &  0.182  & -1.079  &  0.316- & 0.000   &  0.041   & 12.736  &  4.400  & 32.054   & 24.293   & 16.869   \\       
11  &  6.70  & -0.878  & -0.006  &  0.119  &  0.277  & -0.976  &  0.024- & 0.196   &  0.272   & 10.700  &  2.135  &  6.592   & 13.066   &  6.218   \\       
11  &  6.80  & -0.484  &  0.174  &  0.206  &  0.446  & -0.497  &  0.140  & 0.238   &  0.447   &  1.707  &  0.518  &  0.000   &  2.901   &  0.239   \\       
11  &  6.90  & -0.553  &  0.244  &  0.291  &  0.617  & -0.553  &  0.212  & 0.304   &  0.617   &  0.295  &  0.265  &  0.000   &  1.341   &  0.018   \\       
11  &  7.00  & -0.630  &  0.219  &  0.331  &  0.723  & -0.628  &  0.190  & 0.339   &  0.723   &  0.029  &  0.146  &  0.000   &  0.706    & 0.001     \\       
11  &  7.10  & -0.567  &  0.210  &  0.152  &  0.698  & -0.567  & 0.182   & 0.314   &  0.698    & 0.000   & 0.059  &  0.00  &   0.00  &   0.00   \\     
\hline
\end{tabular}
\end{minipage}
\label{contributions}
\end{table*}
 \endlandscape
 \twocolumn



\onecolumn
\landscape

\begin{table*}
\centering
\begin{minipage}{148mm}
\caption{Evolution of broad band and the strongest emission lines
luminosities along the time after each stellar burst occurs.
Only the last burst evolution is shown. The whole table is
available in electronic format.}
 \begin{tabular}{cccccccccccc}
\hline
NB & log $\frac{t}{a}$ & L$_U$ & L$_B$ & L$_V$ & L$_R$ & L$_I$ &
   L([OII]) & L(H$\beta$) & L([OIII]) & L(H$\alpha$) & L([SIII]) \\ &
   & \multicolumn{10}{c}{10$^{40}$ erg s$^{-1}$}\\
11 &   5.75  &  1.559E+01  &  1.423E+01  &  7.080E+00  &  1.123E+01  &  7.442E+00  & 4.252E+00   & 2.474E+00  & 9.574E+00  & 1.1533E+01  & 4.466E+00   \\           
11 &   5.80  &  1.569E+01  &  1.435E+01  &  7.126E+00  &  1.126E+01  &  7.442E+00  & 4.431E+00   & 2.458E+00  & 8.812E+40  & 1.1296E+01  & 4.364E+00   \\           
11 &   5.85  &  1.582E+01  &  1.447E+01  &  7.172E+00  &  1.134E+01  &  7.483E+00  & 4.483E+00   & 2.480E+00  & 9.080E+00  & 1.1560E+01  & 4.459E+00   \\           
11 &   5.90  &  1.595E+01  &  1.460E+01  &  7.225E+00  &  1.139E+01  &  7.510E+00  & 4.716E+00   & 2.487E+00  & 8.731E+00  & 1.1560E+01  & 4.480E+00   \\           
11 &   5.95  &  1.612E+01  &  1.478E+01  &  7.286E+00  &  1.144E+01  &  7.524E+00  & 5.439E+00   & 2.483E+00  & 7.872E+00  & 1.1560E+01  & 4.518E+00   \\           
11 &   6.00  &  1.631E+01  &  1.500E+01  &  7.367E+00  &  1.151E+01  &  7.552E+00  & 6.160E+00   & 2.466E+00  & 7.196E+00  & 1.1560E+01  & 4.538E+00   \\           
11 &   6.05  &  1.654E+01  &  1.519E+01  &  7.448E+00  &  1.161E+01  &  7.601E+00  & 6.303E+00   & 2.476E+00  & 7.016E+00  & 1.1560E+01  & 4.532E+00   \\           
11 &   6.10  &  1.680E+01  &  1.545E+01  &  7.545E+00  &  1.172E+01  &  7.650E+00  & 6.692E+00   & 2.481E+00  & 6.654E+00  & 1.1586E+01  & 4.528E+00   \\           
11 &   6.15  &  1.714E+01  &  1.584E+01  &  7.685E+00  &  1.184E+01  &  7.685E+00  & 7.666E+00   & 2.439E+00  & 5.598E+00  & 1.1349E+01  & 4.439E+00   \\           
11 &   6.20  &  1.751E+01  &  1.622E+01  &  7.835E+00  &  1.199E+01  &  7.756E+00  & 8.309E+00   & 2.425E+00  & 5.082E+00  & 1.1375E+01  & 4.429E+00   \\           
11 &   6.25  &  1.792E+01  &  1.666E+01  &  8.003E+00  &  1.216E+01  &  7.821E+00  & 8.986E+00   & 2.410E+00  & 4.551E+00  & 1.1401E+01  & 4.429E+00   \\           
11 &   6.30  &  1.856E+01  &  1.746E+01  &  8.281E+00  &  1.234E+01  &  7.850E+00  & 1.034E+01   & 2.275E+00  & 2.878E+00  & 1.0689E+01  & 4.158E+00   \\           
11 &   6.35  &  1.927E+01  &  1.833E+01  &  8.599E+00  &  1.257E+01  &  7.908E+00  & 1.065E+01   & 2.143E+00  & 1.871E+00  & 1.0021E+01  & 3.866E+00   \\           
11 &   6.40  &  2.050E+01  &  1.990E+01  &  9.172E+00  &  1.299E+01  &  8.003E+00  & 1.024E+01   & 1.919E+00  & 1.028E+00  & 9.1614E+00  & 3.467E+00   \\           
11 &   6.45  &  2.172E+01  &  2.158E+01  &  9.810E+00  &  1.349E+01  &  8.152E+00  & 8.986E+00   & 1.704E+00  & 6.936E-01  & 7.9977E+00  & 2.946E+00   \\           
11 &   6.50  &  2.478E+01  &  2.554E+01  &  1.136E+01  &  1.485E+01  &  8.631E+00  & 6.600E+00   & 1.282E+00  & 2.535E-01  & 6.0948E+00  & 2.074E+00   \\           
11 &   6.60  &  1.929E+01  &  2.088E+01  &  9.543E+00  &  1.273E+01  &  7.615E+00  & 2.165E+00   & 7.670E-01  & 3.724E+00  & 3.4910E+00  & 1.375E+00   \\           
11 &   6.70  &  1.260E+01  &  1.556E+01  &  7.566E+00  &  1.043E+01  &  6.596E+00  & 1.526E+00   & 3.431E-01  & 5.396E-01  & 1.5847E+00  & 4.419E-01   \\           
11 &   6.80  &  1.012E+01  &  1.795E+01  &  1.023E+01  &  1.528E+01  &  1.041E+01  & 1.776E-01   & 9.445E-02  & 0.000E+00  & 4.6128E-01  & 2.525E-02   \\           
11 &   6.90  &  7.074E+00  &  1.177E+01  &  7.146E+00  &  1.154E+01  &  8.520E+00  & 2.121E-02   & 3.171E-02  & 0.000E+00  & 1.5847E-01  & 1.615E-03   \\           
11 &   7.00  &  5.045E+00  &  7.821E+00  &  4.648E+00  &  7.799E+00  &  6.110E+00  & 1.491E-03   & 1.159E-02  & 0.000E+00  & 5.6100E-02  & 8.127E-05    \\           
11 &   7.10  &  4.390E+00  &  7.205E+00  &  4.247E+00  &  7.009E+00  &  5.456E+00  & 6.406E-08    &4.361E-03   &0.000E+00  & 2.0891E-05  &  0.000E+00     \\         
\hline
\end{tabular}
\end{minipage}
\label{fluxes}
\end{table*}
\endlandscape
\twocolumn

\end{document}